\documentstyle{l-aa}
\begin{document}

  \thesaurus{  11          
              (11.01.1;    
               11.05.2;    
               11.09.4) }  

   \title{On the origin of the luminosity - metallicity relation for
          late-type galaxies}

   \subtitle{Spirals to irregulars transition}

   \author{ L.S. Pilyugin \inst{1},  F.Ferrini \inst{2},}

  \offprints{L.S. Pilyugin }

   \institute{   Main Astronomical Observatory
                 of National Academy of Sciences of Ukraine,
                 Goloseevo, 03680 Kiev-127, Ukraine, \\
                 (pilyugin@mao.kiev.ua)
                  \and
                 Department of Physics, Section of Astronomy,
                 University of Pisa, piazza Torricelli 2,
                 56100 Pisa, Italy,  \\
                 (federico@astr2pi.difi.unipi.it)}
 
   \date{Received ; accepted }

\maketitle

\markboth {L.S.Pilyugin, F.Ferrini: on the origin of the luminosity -
metallicity relation for late-type galaxies}{}

\begin{abstract}

We consider the roles of two widely invoked mechanisms for the 
metallicity-luminosity correlation among late-type galaxies: higher astration 
level and decreasing efficiency of heavy-element loss with increasing 
luminosity. 
We find that both mechanisms contribute about equally to the range in
oxygen abundance between low ($logL_{B} \sim 8$) and high ($logL_{B} \sim 10.5$) 
luminosity galaxies.

We also consider the transition from spirals to irregulars, finding that both 
the oxygen abundance deficiency (indicating the degree of mass exchange
between a galaxy and its environment) and the gas fraction vary smoothly along 
the sequence. Thus we find no "irregular versus spiral dichotomy".


\end{abstract}

\section{Introduction}

Twenty years ago Lequeux et al (1979) have revealed that the oxygen abundance 
correlates with total galaxy mass for irregular galaxies. Since the galaxy
mass is a poorly known parameter, the luminosity - metallicity relation instead 
of the mass - metallicity relation is usually considered (Skillman et al 1989;
Richer \& McCall 1995). It has been found that the characteristic gas-phase 
abundances (the oxygen abundance at a predetermined galactocentric distance) and 
luminosities of spiral galaxies are also correlated (Garnett \& Shields 1987; 
Vila-Costas \& Edmunds 1992; Zaritsky et al 1994, Garnett et al 1997), and 
this relationship maps almost directly on to the luminosity - metallicity 
relationship of irregular galaxies (Zaritsky et al 1994, Garnett et al 1997). 

It is widely suggested that the luminosity - metallicity relation for 
irregular galaxies is caused by galactic winds of different efficiencies. 
In other words, the luminosity - metallicity relation represents the ability 
of a given galaxy to keep the products of its own evolution rather than their 
ability to produce metals (Larson 1974). On the other hand, 
it has been found that the astration level is higher in massive irregular 
galaxies than in dwarf ones (Lequeux et al 1979, Vigroux et al 1987). The
systematic increase of the astration level with luminosity can also play 
a role in the origin of the luminosity - metallicity relationship.

The elucidation of mechanisms which are responsible for a luminosity-
metallicity relation for spirals and irregulars, the clarification of irregulars 
versus spirals dichotomy or the justification of the lack of this dichotomy 
is very important for the understanding of the galaxy formation and evolution.
The values of gas mass fraction $\mu$ and oxygen abundance deficiency (which
is a good indicator of efficiency of mass exchange between a galaxy and
its environments) have been derived for a number of spiral (Pilyugin \&
Ferrini 1998) and irregular (Pilyugin \& Ferrini 2000) galaxies. Using these
data, the roles of two  mechanisms -- astration level increasing with 
luminosity and efficiency of heavy elements loss decreasing -- as causes of the
metallicity - luminosity correlation among later-type spiral and
irregular galaxies will be examined in the present study, along with the 
transition from spirals to irregulars.

\section{The luminosity - metallicity relation for late-type galaxies}

\begin{figure}[thb]
\vspace{7.5cm}
\includegraphics{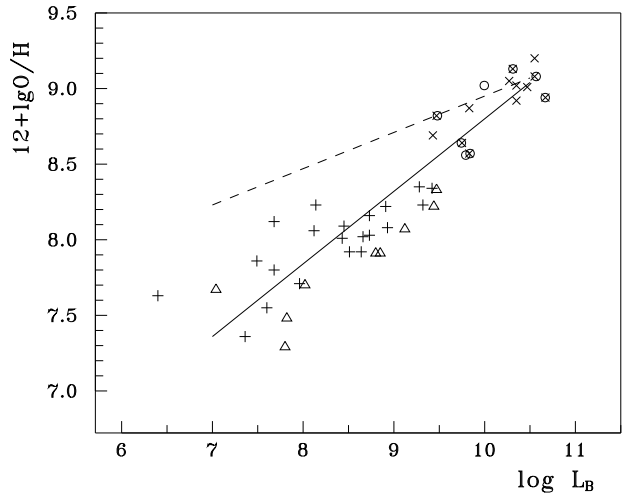}
\caption{\label{figure:9370f1}
$L_{B}$ versus O/H diagram for late-type galaxies. The crosses are late-type 
spirals from Garnett et al (1997). The circles are late-type spirals from
Pilyugin \& Ferrini (1998). The pluses are irregulars from Richer and McCall 
(1995). The triangles are irregulars from Pilyugin \& Ferrini (2000).
The solid line is the adopted relationship between oxygen abundance and 
luminosity. The variation in oxygen abundance with luminosity due to 
the systematic increase of the astration level with luminosity is shown by the 
dashed line (see text). 
}
\end{figure}

The $L_{B}$ versus O/H diagram for late-type galaxies is presented in
Fig.\ref{figure:9370f1}. The positions of late-type spiral galaxies from
Garnett et al (1997) are shown by crosses. The representative oxygen
abundance is the characteristic mean abundance at one disk scale length
from the nucleus. The circles are late-type spiral galaxies for which
the values of global oxygen abundance deficiency $\eta$ and global gas mass 
fraction $\mu$ were derived by Pilyugin \& Ferrini (1998). The positions
of irregular galaxies from Richer \& McCall (1995) are shown with pluses. The
triangles are irregular galaxies for which the values of oxygen abundance 
deficiency and gas mass fraction have been derived by Pilyugin \& Ferrini 
(2000). As can be seen in Fig.\ref{figure:9370f1}, the positions of irregular 
galaxies from our sample are systematically shifted towards high
luminosities 
as compared to the data of Richer and McCall. This small shift is caused by the 
fact that in the work of Richer and McCall (1995) the absolute B magnitudes are 
based on the apparent B magnitudes given in RC3 whereas in ours they
are based on the $B_{T}^{0}$ (total face-on 
magnitude corrected for galactic and internal absorption) given in RC3.
The solid line in Fig.\ref{figure:9370f1} is the adopted relationship 
between oxygen abundance and luminosity. The corresponding equation is
\begin{equation}
12 + \log O/H = 4.0 + 0.48 \times \log L_{B}.    \label{eq:oh-l}
\end{equation}

\begin{figure}[thb]
\vspace{7.5cm}
\includegraphics{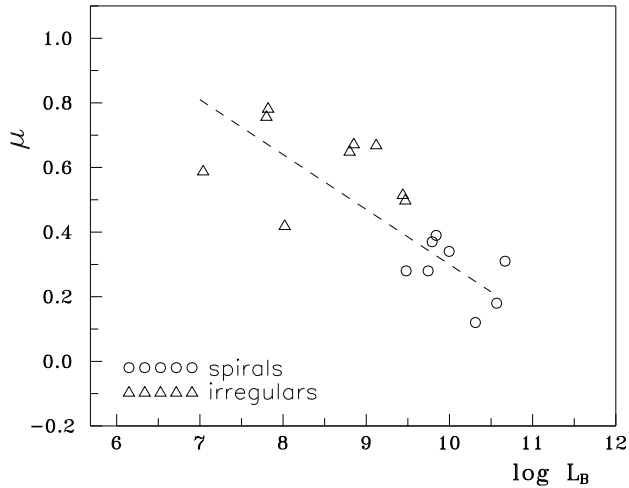}
\caption{\label{figure:9370f2}
$L_{B}$ versus $\mu$ for late-type galaxies. The circles are late-type spirals
from Pilyugin and Ferrini (1998). The triangles are irregulars from Pilyugin
and Ferrini (2000). The line is the adopted relationship between gas mass
fraction  and luminosity. 
}
\end{figure}
 
\begin{figure}[thb]
\vspace{7.5cm}
\includegraphics{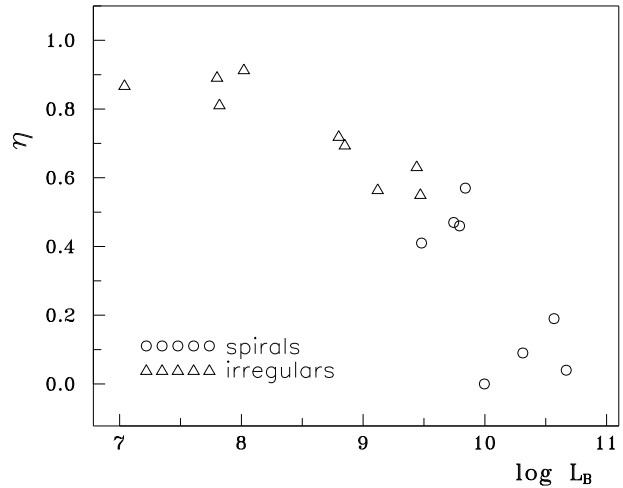}
\caption{\label{figure:9370f3}
$L_{B}$ versus $\eta$  for late-type galaxies. The circles are late-type spirals
from Pilyugin and Ferrini (1998). The triangles are irregulars from Pilyugin
and Ferrini (2000).
}
\end{figure}
 
\begin{figure}[thb]
\vspace{7.5cm}
\includegraphics{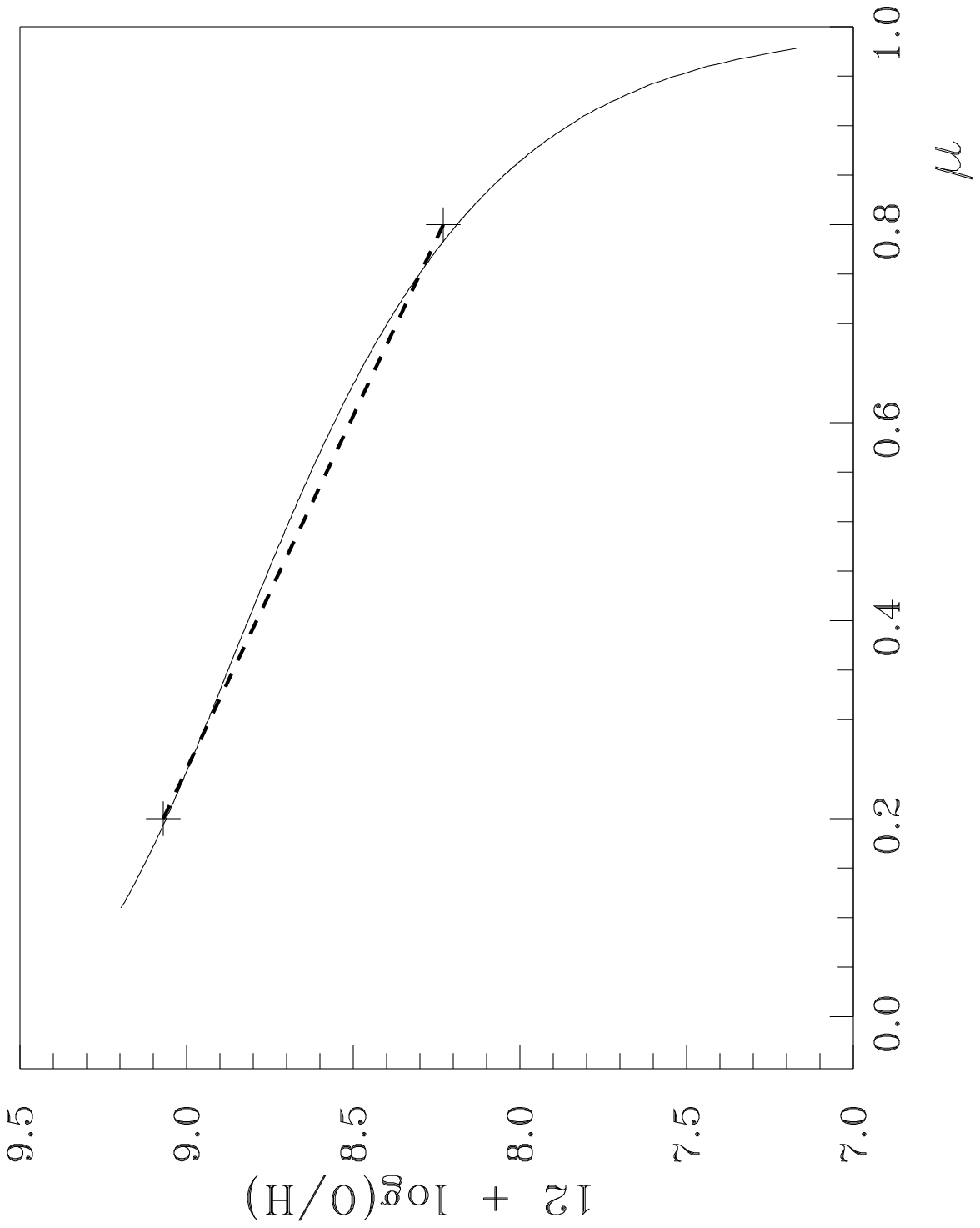}
\caption{\label{figure:9370f4}
$\mu$ versus O/H diagram. 
The positions of one-zone closed-box models with different present-day gas 
mass fraction (the standard curve) is presented by the solid line. 
The dashed line is the linear approximation on the gas mass fraction 
interval from $\mu$ = 0.2 to $\mu$ = 0.8.
}
\end{figure}

The $L_{B}$ versus $\mu$ diagram and the $L_{B}$ versus $\eta$ diagram 
for late-type galaxies are presented in Fig.\ref{figure:9370f2} and
Fig.\ref{figure:9370f3}, respectively. The circles are spiral galaxies from 
Pilyugin \& Ferrini (1998), the triangles are irregular galaxies from 
Pilyugin \& Ferrini (2000).
Inspection of Fig.\ref{figure:9370f2} shows that there is a correlation between
the gas mass fraction $\mu$ and the galaxy's luminosity, decreasing from
$\mu$ $\sim$ 0.8 at log$L_{B}$ = 7 to $\mu$ $\sim$ 0.2 at log$L_{B}$ = 10.5. 
At the same time, as can be seen in the Fig.\ref{figure:9370f3}, the oxygen 
abundance deficiency (which is an indicator of efficiency of galactic winds
or/and the present-day gas infall) increases with decreasing luminosity 
over the large interval of luminosity. Then, one can expect that both
mechanisms contribute to the effect. The available data allow to distinguish
the contribution of the decrease of gas mass fraction with luminosity and the 
contribution of the decrease of galactic wind efficiency with luminosity to 
the difference in oxygen abundances of low- and high-luminosity galaxies.

We discuss first the contribution of the gas mass fraction. The relationship 
between it and galaxy luminosity can be approximated by the linear expression
shown in Fig.\ref{figure:9370f2} by the dashed line.
The corresponding equation is
\begin{equation}
\mu = 2.0 -  0.17 \times \log L_{B}.                \label{eq:mu-l}
\end{equation}
The oxygen abundance as a function of present-day gas fraction has been
computed for closed-box model galaxies using the model of chemical and 
photometric evolution of galaxies by Pilyugin \& Ferrini (1998).
Abundances as a function of present-day gas fraction (standard curve from 
Pilyugin \& Ferrini, 1998, 2000) are shown by the solid curve 
in Fig.\ref{figure:9370f4}. Within the interval $\mu$ = 0.2 to $\mu$ = 0.8 the 
dependence can be well represented by the linear expression 
\begin{equation}
12+\log O/H_{CB} = 9.35 - 1.4 \times \mu  ,          \label{eq:oh-mu-cb}  
\end{equation}
shown with dashed line in Fig.\ref{figure:9370f4}.
The variation in oxygen abundance caused by the increase of 
astration level with luminosity can be derived from Eqs. (\ref{eq:mu-l}) and 
(\ref{eq:oh-mu-cb}), 
\begin{equation}
12+logO/H_{CB} = 6.55 + 0.24 \times logL_{B} ,             \label{eq:oh-l-cb}
\end{equation}
and it is shown in Fig.\ref{figure:9370f1} by a dashed line.
This relation can explain the difference in oxygen abundances between a galaxy 
of luminosity $logL_{B}=8$ and a galaxy of luminosity $logL_{B}=10.5$ around 
0.6 dex, while as it results from Fig.1 the observed difference is around 1.2 
dex. Thus the increase in astration level causes about half of the effect, 
in agreement with the fact that elliptical galaxies have oxygen 
abundances exceeding those of comparably luminous dwarf irregulars (Richer 
\& McCall, 1995), but not all of it.

The deviation of a galaxy's position from the dashed line in
(Fig.\ref{figure:9370f1}) indicates the contribution of heavy elements loss to 
the luminosity -- metallicity correlation. The oxygen abundance $z_{O}$ in the 
interstellar medium of an irregular galaxy evolving with both non selective heavy 
elements loss via ordinary galactic winds and selective heavy elements loss 
via enriched galactic winds is given by 
\begin{equation}
z_{O}  = \frac{p_{O}\;(1 - \lambda_{E}\;f_{O})}
{1 + \lambda/\alpha}\;
ln[\frac{1 + \lambda/\alpha}{\mu} - \frac{\lambda}{\alpha}] ,     \label{eq:zo}
\end{equation}
(Pilyugin 1994), where  $\alpha$ is the proportion of mass in each generation of stars that 
remains locked up in long-lived stars or remnants, $p_{O}$ is the yield of 
oxygen (the mass of new oxygen produced by generation of stars per unit mass 
locked up in long-lived stars or remnants), $\lambda_{O}$ is the efficiency of 
ordinary galactic wind (the ratio of the mass of the ambient interstellar 
medium which leaves the galaxy via galactic wind to the mass of star 
generation which is indirect cause for this wind), $\lambda_{E}$ is the 
efficiency of enriched galactic wind (the fraction of the mass of type II 
supernovae ejecta leaving the galaxy), $f_{O}$ is the contribution of type II 
supernovae to the oxygen production, $\lambda$ is the total efficiency of 
mass loss via ordinary and enriched galactic winds 
\begin{equation}
\lambda = \lambda_{O} + \lambda_{E}\:f_{m}\:(1 - \alpha)  ,   \label{eq:lamda}
\end{equation}
where $f_{m}$ is the contribution of type II supernovae to the cumulative mass 
of matter ejected by a stellar generation into the interstellar medium over the
Hubble time.

In case S of nucleosynthesis (Pilyugin \& Ferrini 1998) we have; 
$p_{O}=0.00866$, $\alpha = 0.206$, $f_{m}= 0.315$, and $f_{O} \approx 1$.
In order to establish the contributions of enriched and ordinary galactic
winds to the total mass loss, the abundances of two elements with different
values of $f_{j}$ should be considered. Since only the oxygen abundance is
considered here, two cases of evolution of a system (evolution with only an
ordinary galactic wind, and evolution with only an enriched galactic wind) 
will be considered.
A "typical" galaxy of luminosity $L_{B}= 10^{8}L_{B,\odot}$ has a gas mass
fraction $\mu = 0.65$ (eq.\ref{eq:mu-l}) and oxygen abundance 12+logO/H =
7.85 (eq.\ref{eq:oh-l}). If this galaxy evolves with enriched galactic
wind only ($\lambda _{O} = 0$) then the efficiency of enriched galactic wind
would be as large as $\lambda _{E} \approx 3/4$ (eq.\ref{eq:zo}).
In other words, a galaxy of luminosity $L_{B}= 10^{8}L_{B,\odot}$  keeps only 
$\sim$ 1/4 of the oxygen produced in course of its evolution. This value is not in 
contradiction with predictions of hydrodynamic models (de Young \& Gallagher 
1990, de Young \& Heckman 1994, McLow \& Ferrara 1999). It should be noted
that the derived efficiency of the enriched galactic wind is an upper limit
since the ordinary galactic wind was not taken into account.
If this galaxy evolves with ordinary galactic wind only ($\lambda _{E} = 0$) 
then the efficiency of ordinary galactic wind would be as large as 
$\lambda _{O} \approx 6.2$. (Again, the derived value of $\lambda _{O}$
is an upper limit since the enriched galactic wind was not taken into account).
In other words, 
a galaxy of luminosity $L_{B}= 10^{8}L_{B,\odot}$  keeps only $\sim 0.1$ of
its initial mass. It should be noted for comparison that a dwarf elliptical 
galaxy can lose of the order of 99 per cent of its initial mass 
(Vigroux et al 1981).

Richer and McCall (1995) have revealed a prominent feature of their
metallicity - luminosity relation for irregular galaxies: they
have found more scatter at low luminosities, though they found less at
high luminosities. The onset of this scatter seems to occur at $M_{B} 
\sim - 15$ or $logL_{B} \sim 8.2$. Moreover, Hidalgo-Gamez and Olofsson 
(1998) have found that there is no relationship between the oxygen
abundance and the absolute magnitude in the blue band for dwarf irregular
galaxies ($M_{B} > - 17$ or $logL_{B} < 9$). 
The following explanation of the disappearance (or increased scatter) of the 
luminosity - metallicity correlation at low luminosities can be suggested,
bearing in mind that the oxygen abundance deficiency becomes constant at low 
luminosities (Fig.\ref{figure:9370f3}). This plateau in the $\eta$ - $L_{B}$ 
relation could be caused by a lack of dependence of galactic winds on 
luminosity at the low-luminosity end, which in turn could result in the absence of 
a relationship between luminosity and  oxygen abundance. However, the dwarf 
galaxies with a well determined oxygen abundance deficiency are few in number, 
so that the edge and even the existence of the plateau are not beyond question. 
 
Thus, both the increase in astration level and the decreasing efficiency
of heavy element loss, with increasing luminosity, make comparable contributions
to the luminosity -- metallicity correlation.


\section{Discussion and conclusions}

By tradition, the late-type galaxies are divided into two classes 
(irregulars and spirals), and these classes of galaxies are investigated 
individually. In particular, the existing models of chemical evolution of spiral 
galaxies are quite different from the ones applied to irregular galaxies. In order to 
reproduce the observational data (mainly for our Galaxy which is a giant spiral) 
various versions of the infall model, in which an infall of gas onto the 
disk takes place for a long time, have been suggested for spiral galaxies 
(Matteucci \& Francois 1989, Ferrini et al 1992, Pardi \& Ferrini 1994,
Pilyugin \& Edmunds 1996, among others). 
The hypothesis of an infall of gas onto the disks of spiral galaxies is 
confirmed by the observational data. Wakker et al (1998) found direct
observational evidence for the infall of low-metallicity gas on the
Milky Way, required in models of galactic chemical evolution.
The Magellanic Stream
is another excellent example of the present-day capture of the matter
by our Galaxy from satellite galaxies.
On the other hand,  various versions of models in which an ejection of gas from the galaxy takes place, have 
been suggested for irregular galaxies (Matteucci \& Chiosi 1983; Matteucci 
\& Tosi 1985; Pilyugin 1993, 1996; Marconi et al 1994; Bradamante et al 
1998, among others).  The hypothesis of gas outflows from dwarf galaxies
is also confirmed by observational data. Outflows of gas 
have been observed in a number of galaxies (Marlowe et al 1995, Heckman 1997). 
However, from Sc to Im the Hubble sequence is really a 
luminosity sequence (Binggeli 1994). The morphological properties of 
galaxies change in a gradual way along a sequence of decreasing luminosity 
(the bulge-to-disk ratio decreases along a sequence of decreasing luminosity 
in a gradual manner, and the bulge is lost in faint galaxies; the spiral 
structure gets more chaotic in fainter galaxies and is also lost) and 
do not show a sharp line of demarcation between irregulars and spirals (an 
"irregular versus spiral dichotomy").

It has been demonstrated (Zaritsky et al 1994, Garnett et al 1997) that
the luminosity - metallicity relationship of spiral galaxies maps almost 
directly on to that of irregulars. This can be considered as evidence that
there is no sharp line of 
demarcation between irregulars and spirals. 
However, spirals have radial oxygen abundance gradients with
slopes in the interval -0.3 $\div$ -0.1 dex per disk scale length
(the slope of the radial oxygen abundance gradient in normal spirals expressed 
in terms of dex/kpc increases, on average, with decreasing luminosity)
(Garnett et al 1997) while irregular galaxies have no radial oxygen abundance 
gradients. The transition from the galaxies with radial oxygen abundance 
gradients to galaxies without them
(or the transition from the galaxies with largest radial oxygen abundance 
gradients in dex/kpc to galaxies without radial oxygen abundance gradients)
takes place at the late end of the Hubble sequence in going from spiral
to irregular galaxies. Thus, the behaviour of oxygen abundance along a 
sequence of decreasing luminosity is twofold. On one hand, the luminosity - 
metallicity relationship seems to be unique both for spirals and for irregulars. 
On the other hand, a jump-like change of slope of radial oxygen abundance 
gradient takes place in going from spiral to irregular galaxies.

The unique  luminosity - metallicity relationship for late-type spiral
and irregular galaxies is not indisputable since this relationship for
spiral galaxies depends on the choice of representative oxygen abundance.
In investigations of the relationship between the oxygen abundance and the 
luminosity of spiral galaxies, the concept of the characteristic oxygen 
abundance  has been introduced: it is defined as the oxygen abundance at a
predetermined galactocentric distance $r^{*}$. Owing to the presence of radial
abundance gradients, the choice of the "representative" abundance for spirals
is not a trivial matter. Choices of $r^{*}$ have included zero (giving 
the extrapolated central intersect abundance), $r^{*} = 0.4 \rho _{0}$ 
(where $\rho _{0}$ is the isophotal radius with surface brightness 25 mag 
arcsec$^{-2}$) and the disk scale length (Vila-Costas \& Edmunds 1992; 
Zaritsky, Kennicutt \& Huchra 1994; Garnett et al 1997).  With any choice of 
$r^{*}$, the value of  $O/H(r^{*})$ is local parameter, i.e. it represents the 
oxygen abundance in the region at a given galactocentric distance but not 
the oxygen abundance of the whole galaxy.

The $L_{B}$ versus $\mu$ diagram (Fig.\ref{figure:9370f2}) and the $L_{B}$ 
versus $\eta$ diagram (Fig.\ref{figure:9370f3}) provides an additional test
of whether the properties of late-type galaxies change in a gradual way along 
a sequence of decreasing luminosity. 
Inspection of Figs.\ref{figure:9370f2} and \ref{figure:9370f3} shows that there are 
no jump-like variations in astration level and oxygen abundance 
deficiency in going from spiral to irregular galaxies. 
Thus, the variations of astration level and oxygen abundance deficiency among
late-type spiral and irregular galaxies provide additional evidence 
that the properties of late-type galaxies change in a gradual way along 
a sequence of decreasing luminosity.

The jump-like change of the value of the slope of the radial oxygen abundance
gradient in going from spiral to 
irregular galaxies may be attributed to the following effect. It is known
that the barred spiral galaxies have no or have a smaller values of the
radial oxygen abundance
gradients as compared to the non-barred spirals (Martin \& Roy 1994, Friedli 
1999). Almost all  Sm galaxies are barred galaxies (Odewahn 1996, 
Friedli 1999), and therefore they have no radial oxygen abundance gradients.
Thus, the jump-like change in the average 
gradient in going from spirals to irregulars seems to be caused by a 
variation in the fraction of barred galaxies.

There is a number of still open questions. Is there a sharp line of
demarcation between the galaxies evolving with gas infall to galaxies evolving 
with mass ejection? Or do these kinds of mass exchange between the galaxy and its 
environment take place simultaneously in some galaxies? What kind of the 
chemical evolution model should be applied to low-luminous spirals of type Sd?

\begin{acknowledgements}
We thank Dr. N.Bergvall for his constructive comments on the manuscript.
We thank the referee, Prof. B.E.J.Pagel, for helpful comments and suggestions
which resulted in a better presentation of the work. 
L.P. thanks the Staff of Department of Physics, Section of Astronomy 
(University of Pisa) for hospitality. This study was partly supported by the 
INTAS grant No 97-0033. 
\end{acknowledgements}

\end{document}